\RequirePackage{lineno}
\documentclass[preprint,preprintnumbers]{revtex4}
\usepackage{graphicx}
\usepackage{dcolumn}
\usepackage{bm}

\begin{document}

\title{Atom-light superposition oscillation and Ramsey-like atom-light
interferometer}
\author{Cheng Qiu$^1$, Shuying Chen$^1$, L. Q. Chen$^{1,*}$, Bing Chen$^1$,
Jinxian Guo$^1$, Z. Y. Ou$^{1,2,\dag}$, and Weiping Zhang$^{1,\ddag}$}
\affiliation{$^{1}$Department of Physics, East China Normal University, Shanghai 200062,
P. R. China\\
$^{2}$Department of Physics, Indiana University-Purdue University
Indianapolis, 402 North Blackford Street, Indianapolis, Indiana 46202, USA}
\maketitle

\noindent\textbf{Coherent wave splitting is crucial in interferometers.
Normally, the waves after this splitting are of the same type. But recent
progress in interaction between atom and light has led to the coherent
conversion of photon to atomic excitation. This makes it possible to split
an incoming light wave into a coherent superposition state of atom and light
and paves the way for an interferometer made of different types of waves.
Here we report on a Rabi-like coherent-superposition oscillation observed
between atom and light and a coherent mixing of light wave with excited
atomic spin wave in a Raman process. We construct a new kind of hybrid
interferometer based on the atom-light coherent superposition state.
Interference fringes are observed in both optical output intensity and
atomic output in terms of the atomic spin wave strength when we scan either
or both of the optical and atomic phases. Such a hybrid interferometer can
be used to interrogate atomic states by optical detection and will find its
applications in precision measurement and quantum control of atoms and light.%
}

%\linenumbers
%\begin{linenumbers}

In quantum storage, complete conversion of quantum states between atoms and
light is essential for the high fidelity transfer of quantum information.
Quantum storage was first realized with the method of electromagnetically
induced transparency \cite{fle,hau,wal1,wal2,squ1,squ2}. More recently,
Raman processes were used to achieve wide band quantum storage \cite%
{wam0,wam1,wam2}. On the one hand, most researches concentrated on
increasing the efficiency of quantum storage because a partial conversion is
usually regarded as a loss for the quantum system and leads to a reduction
in the fidelity of quantum information transfer. But the unconverted part
still contains the original information. So, if available, it can be further
converted \cite{wam3} for better overall conversion efficiency. On the other
hand, since quantum storage is coherent in the sense that the phase of the
quantum states is preserved, the converted and the unconverted parts are
coherent to each other. This property can be employed for quantum
interference. Indeed, Campell et al \cite{cam} achieved coherent mixture of
atomic wave and optical wave in a atom-photon polariton state with gradient
echo memory scheme.

Rabi oscillation is a coherent population oscillation between two atomic
levels when driven by a strong coherent radiation field coupled to the two
levels \cite{eb}. It played an important role in atomic clocks by forming a
Ramsey atomic interferometer \cite{ram}. Two-photon Rabi oscillation was
also realized in an atomic Raman system where two strong driving fields are
present \cite{2rabi}. Recently, Rabi oscillation between photons of Raman
write field and the frequency-offset Stokes field was demonstrated \cite%
{chen10} in Raman process where the driving field is a strong atomic spin
wave. Here, the roles were reversed for atom and light as compared to the
traditional Rabi oscillation effect. It was recently predicted \cite{ou08}
that Rabi-like coherent-superposition oscillation between light and atom can
also occur in an atomic Raman process. When the driving field is a $\pi $%
-pulse, it is possible to make a complete conversion from light to atom for
quantum storage or from atom to light for readout.

However, when the driving field is a $\pi /2$-pulse, we can achieve a
coherent wave splitting of the input field into an optical wave and an
atomic wave. The reverse process is just a coherent mixing of an optical
wave and an atomic spin wave. Thus, it is possible to form a new type of
interferometer made of atom and light. In contrast to the traditional
interferometers, which are constructed with linear beam splitters for
coherent splitting into a mixing of the same type of wave and are only
sensitive to the phase shift of one type of wave, this new hybrid atom-light
interferometer involves waves of different type and should depend on the
phases of both optical and atomic waves. This is somewhat similar to an
SU(1,1) type atom-light interferometer recently realized in our group \cite%
{chen2015}

In this paper, we report on the first observation of Rabi-like oscillation
between light and atom in a Raman process involving Rb-atoms and a
demonstration of an atom-light interferometer by employing this Rabi-like
oscillation effect as atom and light wave splitter and mixer. This is a
Ramsey-type interferometer in the sense that a strong driving laser in $\pi
/2$-pulse area creates a superposition between atom and light and after a
time delay with the evolution of both atom and light, another $\pi /2$-pulse
laser is applied to mix atom and light for interference. In additional to
the usual dependence on the optical phase, we find that the interference
fringes also depend on the atomic phase, which is sensitive to a variety of
physical quantities. Thus, this type of interferometer can be applied in
precision measurement, sensitive measurement of atomic states, and quantum
control of light and atoms.

\begin{figure}[htbp]
\includegraphics[width=3.5in]{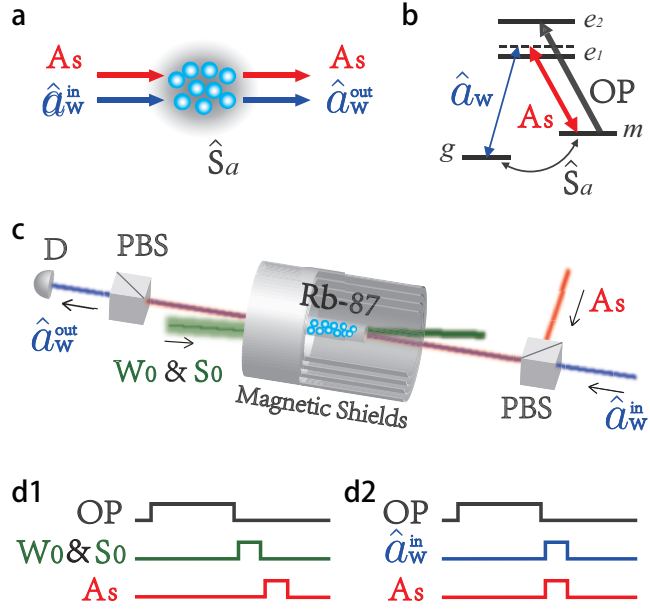}
\caption{\textbf{Experimental set-up of} \textbf{Rabi-like coherent
oscillation.} \textbf{a.} Raman process for three-wave coupling of two
optical fields (write $\hat{a}_{W}$ and Stokes $A_{S}$) and an atomic spin
wave($\hat{S}_{a}$). \textbf{b}. Atomic energy levels for coupling optical
fields. $|g\rangle $ and $|m\rangle $: two ground states $\left\vert
5^{2}S_{1/2},F=1,2\right\rangle $; $|e_{1}\rangle $ and $|e_{2}\rangle $:
two excited states $\left\vert 5^{2}P_{1/2},F=2\right\rangle $ and $%
\left\vert 5^{2}P_{3/2}\right\rangle $. OP: optical pumping field resonant
on the $|m\rangle \rightarrow |e_{2}\rangle $ transition. \textbf{c}.
Experimental arrangement for observing Rabi oscillation between atom and
light driven by a strong Stokes field with the atoms having an initial spin
wave. The initial spin wave is prepared by W$_{0}$\&S$_{0}$ (see Method for
detail). PBS: polarized beam splitter. \textbf{d1} and \textbf{d2} Timing
sequences for the experiment with an initial spin wave $\hat{S}_{a}^{in}$
(d1) and a write field $\hat{a}_{W}^{in}$ (d2) as the only input field. }
\end{figure}

The process we use to mix atomic and optical waves is the collective Raman
process in an ensemble of $N_{a}$ three-level atoms. The process is depicted
in Fig.1a with the atomic levels and optical frequencies shown in Fig.1b. In
the process, a pair of lower level meta-stable states $|g\rangle ,|m\rangle $
is coupled to the Raman write field (W, or $\hat{a}_{W}$) and the Stokes
field (S, or $\hat{a}_{S}$) via an upper excited level $e$. After
adiabatically eliminating the upper excited level $e$, this process is a
three-wave mixing process involving the write field, the Stokes field and a
collective atomic pseudo-spin field $\hat{S}_{a}\equiv (1/\sqrt{N_{a}}%
)\sum_{k}|g\rangle _{k}\langle m|$ and the coupling Hamiltonian is given by 
\cite{DLCZ,pol}: 
\begin{equation}
\hat{H}_{R}=i\hbar \eta \left( \hat{a}_{W}\hat{a}_{S}^{\dag }\hat{S}%
_{a}^{\dag }-\hat{a}_{W}^{\dag }\hat{a}_{S}\hat{S}_{a}\right) ,  \label{Hr}
\end{equation}%
where $\eta =g_{eg}g_{em}/\Delta $ with $g_{eg},g_{em}$ as the coupling
coefficients between the excited state and the lower level states. $\Delta $
is the detuning from the excited state for both the Stokes and Raman pump
fields, which satisfy the two-photon resonance condition: $\omega
_{W}-\omega _{S}=\omega _{mg}$. When the Stokes field $\hat{a}_{S}$ is very
strong coherent field, replacing it with a c-number $A_{S}$ in Eq.(\ref{Hr})
and defining a Rabi-like frequency $\Omega =2\eta A_{S}^{\ast }$, we have
the Hamiltonian 
\begin{equation}
\hat{H}_{BS}^{AL}=\frac{1}{2}i\hbar \left( \Omega ^{\ast }\hat{a}_{W}\hat{S}%
_{a}^{\dag }-\Omega \hat{a}_{W}^{\dag }\hat{S}_{a}\right) .
\end{equation}%
This Hamiltonian leads to the time evolution of the fields given by 
\begin{eqnarray}
\hat{a}_{W}^{out} &=&\hat{a}_{W}^{in}\cos \left( \theta /2\right) +\hat{S}%
_{a}^{in}\sin \left( \theta /2\right) ,  \label{opRm} \\
\hat{S}_{a}^{out} &=&\hat{S}_{a}^{in}\cos \left( \theta /2\right) -\hat{a}%
_{W}^{in}\sin \left( \theta /2\right) ,  \nonumber
\end{eqnarray}%
where $\theta $ is equal to $|\Omega |t$, with $t$ the evolution time.

It is interesting to see that if there is only one input field, say the
write field ($I_{W}^{(0)}\neq 0$), intensities of the output fields are $%
I_{W}=I_{W}^{(0)}\cos ^{2}\left( \theta /2\right) ,I_{S_{a}}=I_{W}^{(0)}\sin
^{2}\left( \theta /2\right) $, respectively, which oscillate in time with a
frequency proportional to $A_{S}$, the amplitude of the strong Stokes field.
This resembles the Rabi oscillation in a two level system driven by a strong
field \cite{eb}. Here, the oscillation is between the write field and the
atomic spin wave instead of the atomic levels while the strong Stokes field (%
$A_{S}$) is the driving field. Similar oscillation occurs if only the atomic
spin wave is initially non-zero ($I_{S_{a}}^{(0)}\neq 0$): $%
I_{W}=I_{S_{a}}^{(0)}\sin ^{2}\left( \theta /2\right)
,I_{S_{a}}=I_{S_{a}}^{(0)}\cos ^{2}\left( \theta /2\right) $. On the other
hand, if both fields are initially non-zero, the outputs are coherent
mixture of the two fields with mixing coefficients as $\sin (\theta /2)$ and 
$\cos (\theta /2)$.

\begin{figure}[htbp]
\includegraphics[width=3.3in]{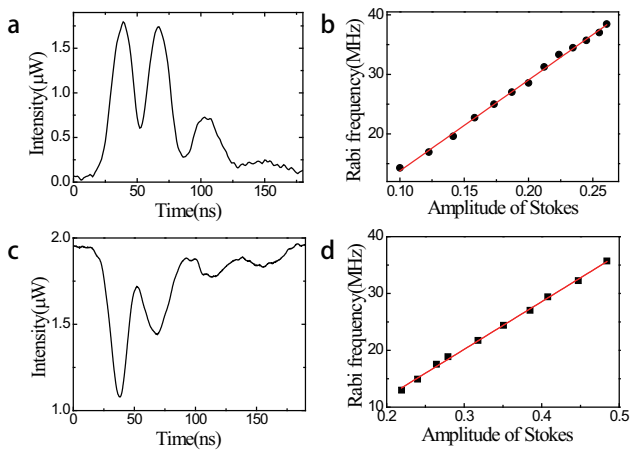}
\caption{\textbf{Results of Rabi-like oscillation between atom and light.} 
\textbf{a}. The output write field observed in time when the atomic spin
wave has an initial value but no write field injection and \textbf{b.} the
corresponding oscillation frequency as a function of the amplitude of the
driving Stokes field. \textbf{c.} The observed output write field when the
write field has an input but the atoms are all in the ground state and 
\textbf{d.} the corresponding Rabi oscillation frequency as a function of
the amplitude of the driving Stokes field.}
\end{figure}

For the experimental observation of the Rabi oscillation between light and
atom, we can approach by either preparing the atoms with an initial atomic
spin wave or simply injecting a write field. The experimental sketch is
shown in Fig.1(c) with timing sequence in Fig.1(d1). In the first approach,
atoms are initially prepared with a non-zero spin wave by two pulses of S$%
_{0}$ and W$_{0}$ (see Method for detail). After a short delay, a strong
Stokes driving pulse ($S$) of 0.2 $\mu $s length is sent in opposite
direction into the cell to drive the Rabi oscillation. To observe it,
W-field ($\hat{a}_{W}^{out}$) is measured by a photo-detector and recorded
in an oscilloscope. Fig.2(a) shows a typical run, clearly demonstrating the
oscillation effect. The amplitude decay is due to the decoherence of the
atomic spin wave. Fig.2(b) shows the oscillation frequency as a function of
the amplitude $A_{S}$ of the strong driving Stokes field. The linear
dependence confirms it as a Rabi frequency, as given in Eq.(\ref{opRm}). For
the case of initial injection at the W-field ($\hat{a}_{W}^{in}$ shown in
Fig.1), we need to lock its frequency to the strong Stokes driving field to
within several hundred Hertz to satisfy the requirement of the two-photon
resonance condition: $\omega _{W}-\omega _{S}=\omega _{gm}$. Time sequence
is shown in Fig.1(d2). Fig.2(c) shows a typical run for this case. The
oscillation frequency is confirmed in Fig.2(d) as Rabi frequency given in
Eq.(\ref{opRm}). Notice that the curves in Fig.2(a) and (c) are
complementary to each other, reflecting the sine and cosine functions in
Eq.(3).

\begin{figure}[htbp]
\includegraphics[width=6.0in]{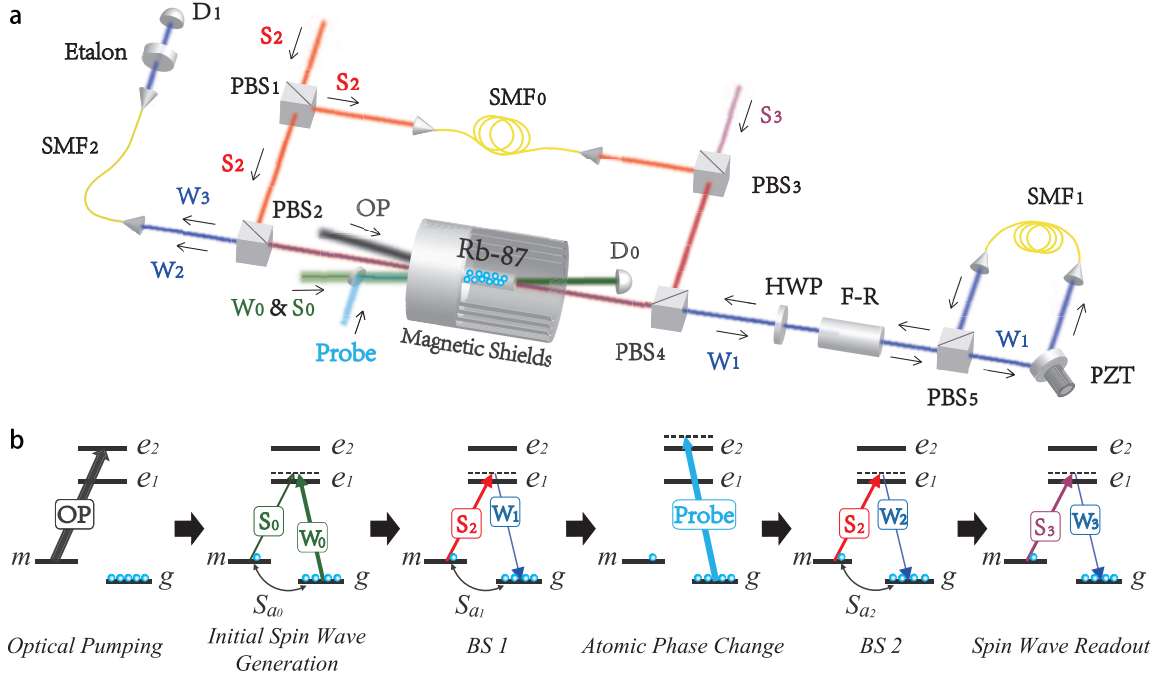}
\caption{\textbf{Atom-light hybird interferometer. a.} Experimental set-up
and \textbf{b.} atomic energy levels with frequencies of optical fields for
the formation of an atom-light hybrid interferometer. HWP: half wave plate;
F-R: Faraday rotator. PZT: Piezoe-lectric Transducer; SMF: single-mode
fiber. W$_{1}$, W$_{2}$, W$_{3}$: write fields; S$_{2}$, S$_{3}$: strong
Stokes fields. BS1: the first splitting process to split the initial spin
wave $S_{a0}$ to coherent superposition of write field W$_{1}$ and spin wave 
$S_{a1}$. BS2: the second beam splitting process to mix W$_{1}$\ and $S_{a1}$
and output W$_{2}$\ and $S_{a2}$.}
\end{figure}

\begin{figure}[htbp]
\includegraphics[width=3in]{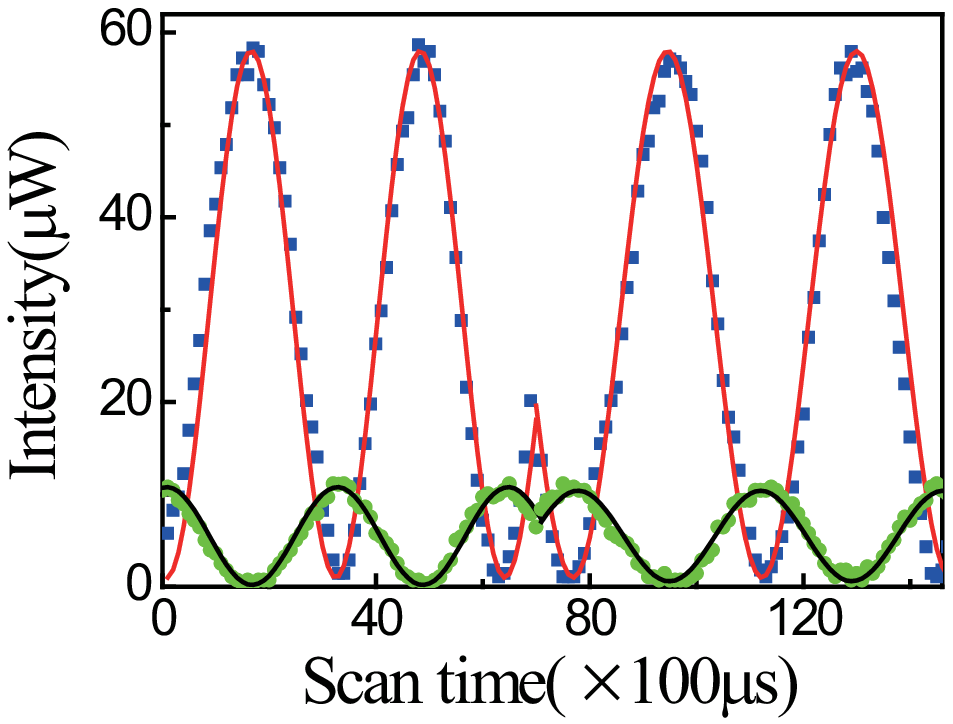}
\caption{\textbf{Interference fringes of atom-light hybrid interferometer. }%
Observed interference fringes at the output write field (blue squares) and
for the final atomic spin wave (green dots) as the optical phase is scanned
via a ramp voltage on the PZT.}
\end{figure}

From Eq.(\ref{opRm}), we find when we adjust the pulse width or amplitude of
the Stokes field so that $\theta =\pi $, the oscillation stops at the
maximum conversion between atom and light. Notice that the input-output
relation in Eq.(\ref{opRm}) is exactly the relation for a lossless beam
splitter \cite{ou87,teich90}: the write field here is equivalent to one of
the input fields of the beam splitter and the atomic spin wave is the other
field. Thus the outputs are coherent mixtures of the optical field and the
atomic spin wave. When the pulse width satisfies $\theta =\pi /2$, only half
will be converted and this leads to a coherent atom-light wave splitting.

Next, we use this atom-light wave splitter to form an atom-light
interferometer. As shown in Fig.3, after the first splitting of the incoming
wave (the initially prepared atomic spin wave here in our experiment), we
mix the split atom and light waves with another but similar conversion
process. This is done by redirecting the generated W-field (W$_{1}$) back to
the atomic cell which contains the unconverted atomic spin wave and mixing
them with another Stokes $\pi /2$-pulse. To separate the splitting and the
mixing processes, a delay between the two Stokes $\pi /2$ pulses is
introduced with a 100-meter-long single mode fiber (SMF$_{0}$). Similar
delay with another SMF$_{1}$ is introduced in the returned W-field.

To make a comparison with a conventional interferometer with beam splitters,
the first pulse will act as the Raman read field in the first Raman read
process to split the initially prepared atomic spin wave $S_{a0}$ into half
write field (W$_{1}$) and half spin wave ($S_{a1}$). This process can be
regarded as the first beam splitter in the conventional interferometer. The
lengths of the two SMFs are made equal to ensure that the delayed Stokes
pulse (S$_{2}$) and the write field W$_{1}$, generated in the first Raman
process and fed-back to the cell, will enter the vapor cell that contains
the spin wave $S_{a1}$ at the same time to carry out the second Raman
process for mixing $S_{a1}$ and W$_{1}$, i.e., the second beam splitter
process. In between the first splitting and the second mixing processes, we
introduce a phase modulation unit (PZT) in W$_{1}$'s path to change its
phase. The final light signal in the write field (W$_{2}$) after the second
BS is collected with another single mode fiber (SMF$_{2}$) and detected
after an etalon, which is used to filter out the leaked strong Stokes
photons. The atomic signal after the second BS can be converted into the
light field (W$_{3}$) by injecting another strong read pulse (S$_{3}$) right
after the second Stokes pulse (S$_{2}$). The conversion efficiency is about
20\%. This signal is also collected by SMF$_{2}$ and detected after the
etalon. So we will observe two temporally separated pulses by the detector:
first one from W$_{2}$ and second one from W$_{3}$. The heights of the two
pulses correspond to the intensities of the final write field (W$_{2}$) and
the final atomic spin wave $S_{a2}$, respectively. Fig.4 shows the
interference fringes detected in both the output W-field and the final
atomic spin wave. The solid lines are the best fits to the cosine function
with visibilities of 96.6$\%$ and 94.8$\%$, respectively.

\begin{figure}[htbp]
\includegraphics[width=3in]{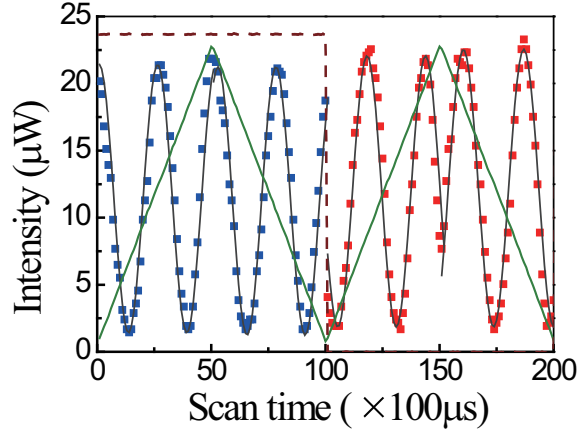}
\caption{\textbf{AC Stark effect on interference output.} Interference
fringes at the output write field with (blue) and without (red) the atomic
phase shift induced by the AC Stark effect. Green lines are the ramp voltage
on the PZT for phase scan and the dotted lines are for probe light intensity
(scales are not drawn for both lines)}
\end{figure}

\begin{figure}[htbp]
\includegraphics[width=4.5in]{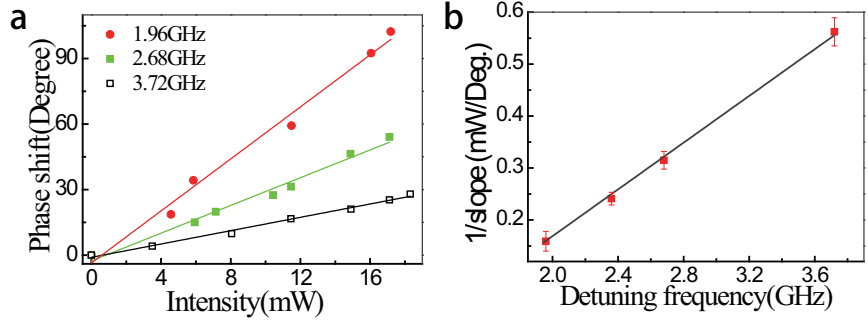}
\caption{\textbf{Atomic phase shift.} \textbf{a.} The induced phase shift as
a function of the intensity of the inducing light field (probe). \textbf{b.}
The inverse of the slopes of the linear fits from (a) as a function of
detuning frequency of probe field. }
\end{figure}

Since the atomic spin wave is involved in this interference scheme, the
interference fringes should depend on the phase of the atomic spin wave as
well, which can be changed by some external field, such as magnetic field
and electric field. However, dependence on the magnetic field relies on the
magnetic sub-levels and can be very complicated. Here, we resort to an AC
Stark effect \cite{AC} for the atomic phase change. It is well-known that
when atoms are subject to the illumination of an electromagnetic field,
their energy level will be shifted. For the atomic Raman process discussed
previously, the AC Stark shift is given by the amount of \cite{pol} 
\begin{equation}
\Delta \Omega _{AC}=g_{eg}g_{em}|E|^{2}/\Delta ,  \label{sta}
\end{equation}%
where $E$ is the amplitude of the field and $\Delta $ is the detuning. This
will lead to a phase shift of $\Delta \varphi =\Delta \Omega _{AC}\Delta T$
for an interaction time of $\Delta T$. Thus, the atomic phase is directly
proportional to the intensity $I\propto |E|^{2}$ of the applied field and
the optical delay $\Delta T$. In the experiment, the external field (called
as probe) comes from another laser at 780 nm tuned to D2 line with 2 to 3
GHz detuning. We turn on the laser after the first Raman splitting process
for $\Delta T=80$ ns and turned it off right before the second Raman mixing
process. In this way, the phase-shifting field will not affect the Raman
processes for splitting and mixing the atomic and optical waves. Fig.5 shows
two interference fringes with and without the 780nm-laser turned on. The
laser power is about 45 mW. We can extract a phase shift of 2.5 radian from
the two fringes. In Fig.6(a), we plot the phase shift as a function of the
intensity of the 780nm-laser at various detuning. Then the inverse of the
slopes of the linear fits from Fig.6(a) are plot against the detuning. The
linear dependence shown in both Fig.6(a) and (b) is in agreement with Eq.(%
\ref{sta}). From Fig.6, we find $\Delta \varphi =\kappa P\Delta T/\Delta $
with $\kappa =0.06Degree\cdot GHz/ns\cdot mW$ for $\Delta T=80$ ns.

In summary, we demonstrate coherent conversion between atom and light in the
form of a Rabi-like oscillation. This coherent conversion process can be
used as a wave splitter into coherent superposition of atomic spin wave and
optical wave and as a wave mixer of the coherent atomic spin wave and
optical wave. We construct an atom-light hybrid interferometer in which the
interference fringes depend on both the optical phase and atomic phase. The
intensity-dependent atomic phase shift can be used for a quantum
non-demolition measurement (QND) \cite{bra,Lap,per} of the photon number of
the phase-inducing field (the 780nm-laser beam in our experiment). This is
similar to the QND measurement in optical Kerr effect \cite{lev,fri,poi}.
Furthermore, atomic phase can be changed by other means such as magnetic and
electric field. So, this atom-light interferometer will have wide
applications in precision measurement, quantum metrology, and quantum
control of atom and light.

%\end{linenumbers}

\noindent\textbf{Method}

In the experiment, the atomic medium is Rubidium-87 atoms which is contained
in a 50mm long paraffin coated glass cell. The cell is placed inside a
four-layer magnetic shielding to reduce stray magnetic fields and is heated
up to 75$^{\circ }$ using a bi-filar resistive heater. The energy levels of
the Rb atom are shown in Fig.3b, where states $|g\rangle $ and $|m\rangle $
are the two ground states ($5^{2}S_{1/2}F=1,2$) from hyperfine splitting and 
$|e_{1}\rangle ,|e_{2}\rangle $ are two excited states ($%
5^{2}P_{1/2},5^{2}P_{3/2}$). An optical pumping field (OP), tuned to $%
|m\rangle \rightarrow |e_{2}\rangle $ transition at 780 nm, is used to
prepare the atoms in $|g\rangle $ state. To get the initial atomic spin wave 
$S_{a0}$, we apply the Raman write field (W$_{0}$) and Stokes seed ($a_{S0}$%
) simultaneously to perform a stimulated Raman scattering. The write laser
is blue-detuned about 1.5 GHz from $|g\rangle \rightarrow |e_{1}\rangle $
transition. The Stokes seed we used comes from the same laser as the W$_{0}$
beam but its frequency is red-shifted 6.8 GHz from W$_{0}$ beam in another
vapor cell using feedback Raman scattering \cite{chen13}.

*lqchen@phy.ecnu.edu.cn

{\dag}zou@iupui.edu

{\ddag}wpzhang@phy.ecnu.edu.cn

\noindent\textbf{Acknowledgements}

This work was supported by the National Basic Research Program of China (973
Program) under Grant No. 2011CB921604, the National Natural Science
Foundation of China (Grant No. 11274118, 91536114, 11129402 and 11234003)
and Supported by Innovation Program of Shanghai Municipal Education
Commission (Grant No. 13ZZ036).

\noindent\textbf{Author contributions}

CQ, SC, BC and JG performed the experiment under the supervision of LQC and
ZYO. CQ, SC, LQC analyzed the data. LQC, ZYO and WPZ wrote the paper. WPZ
provides the overall supervision on this project.

\end{document}